\documentclass[twocolumn,seceq]{jpsj2} 

\def\kf{k_{\rm F}}

\def\ggs{\buildrel\textstyle > \over {\hbox{\raise0.2ex\hbox{$\sim$}}}}
\def\lls{\buildrel\textstyle < \over {\hbox{\raise0.2ex\hbox{$\sim$}}}}
\def\gsim{\,\lower0.75ex\hbox{$\ggs$}\,}
\def\lsim{\,\lower0.75ex\hbox{$\lls$}\,}

\def\et{{\it et al.}}

\def\jo #1#2#3#4{#1 {\bf #2} (#3) #4}   
\def\PR{Phys.\ Rev.}
\def\PRB{Phys.\ Rev.\ B}
\def\PRL{Phys.\ Rev.\ Lett.}

\def\JPSJ{J.\ Phys.\ Soc.\ Jpn.}

\def\CR{Chem.\ Rev.} 

\title{Charge Fluctuations in Geometrically Frustrated Charge Ordering System} 

\author{Hitoshi \textsc{Seo}$^{1,2}$\thanks{email address: seo0@spring8.or.jp}, 
Kenji \textsc{Tsutsui}$^{3}$, 
Masao \textsc{Ogata}$^{4}$, 
and Jaime \textsc{Merino}$^{5}$
}
\inst{$^{1}$Non-Equilibrium Dynamics Project, ERATO-JST, 
c/o KEK, Tsukuba 305-0801, Japan\\
$^{2}$Japan Atomic Energy Research Institute (JAEA), c/o SPring-8, Hyogo 679-5148, Japan\\
$^{3}$Institute for Material Reseach, Tohoku University, Sendai 980-8577, Japan\\
$^{4}$Department of Physics, Faculty of Science, University of Tokyo,
Tokyo 113-0033, Japan\\
$^{5}$Departamento de F\'isica Te\'orica de la Materia 
Condensada, Universidad Aut\'onoma de Madrid, Madrid 28049, Spain
}

\abst{Effects of geometrical frustration 
in low-dimensional charge ordering systems are theoretically studied, 
mainly focusing on dynamical properties. 
We treat extended Hubbard models at quarter-filling, 
where the frustration arises from competing charge ordered patterns 
favored by different intersite Coulomb interactions, 
which are effective models for 
various charge transfer-type molecular conductors and transition metal oxides. 
Two different lattice structures are considered: 
(a) one-dimensional chain with intersite Coulomb interaction of 
nearest neighbor $V_1$ and that of next-nearest neighbor $V_2$, and 
(b) two-dimensional square lattice with $V_1$ along the squares 
and $V_2$ along one of the diagonals. 
From previous studies, 
charge ordered insulating states are known to be unstable 
in the frustrated region, i.e., $V_1 \simeq 2V_2$ for case (a) and  $V_1 \simeq V_2$ for case (b), 
resulting in a robust metallic phase 
even when the interaction strenghs are strong. 
By applying the Lanczos exact diagonalization to finite-size clusters, 
we have found that fluctuations of different charge order patterns exist 
in the frustration-induced metallic phase, 
showing up as characteristic low energy modes
in dynamical correlation functions. 
Comparison of such features between the two models are discussed, 
whose difference will be ascribed to the dimensionality effect. 
We also point out incommensurate correlation in the charge sector 
due to the frustration, 
found in one-dimensional clusters. 
}

\kword{charge order, geometrical frustration, molecular conductors, transition metal oxides, 
exact diagonalization, extended Hubbard model}

\begin{document}
\maketitle
\section{Introduction}\label{sec_intro}

Charge ordering (CO) phenomena have attracted interest from early days, 
motivated by experiments on various materials. 
A classic example is transition metal oxide Fe$_3$O$_4$, the magnetite, 
where its metal-insulator transition 
was proposed by Verwey to be due to CO among the electrons of the Fe ions 
at the so-called $B$-sites~\cite{Verwey39Nature,Fe3O4Review}. 
However, this issue has not been solved 
despite enormous amount of works, and even the existence of CO itself 
has recently been doubted~\cite{Novak00PRB,Garcia00PRB,Todo01JAP,Seo02PRB}. 
A reason for such puzzle is that this CO system 
is under strong geometrical frustration as pointed out by Anderson~\cite{Anderson56PR}. 
This is seen when the system is described by 
Ising variables $\sigma=\pm 1$ corresponding to occupied/unoccupied sites, 
on the pyroclore lattice formed by the $B$-sites. 
An antiferromagnetic interaction 
for each nearest neighbor $\langle ij \rangle$ bond
arises due to the Coulomb repulsion favoring $\sigma_i \sigma_j = -1$~\cite{Anderson56PR}. 
This is indeed a typical situation of geometrical frustration, 
as has frequently been discussed in spin systems~\cite{FrustrationReview}. 

CO systems under geometrical frustration have renewed its interest 
rather recently, triggered by different experimental works. 
The one-dimensional (1D) Cu-O unit in a transition metal oxide PrBa$_2$Cu$_4$O$_8$~\cite{Horii00PRB}
has been pointed out to be susceptible to frustration 
due to its zigzag chain structure~\cite{Seo01PRB}. 
A class of molecular conductors, 
i.e., $\theta$-ET$_2X$, where ET is an abbreviation for 
BEDT-TTF (bisethilenodithio-tetrathiofulvalene) molecule 
and $X$ stands for different monovalent anions with closed shell~\cite{HMori98PRB}, 
is a clear example of systems affected by frustration in their two-dimensional (2D) layer~\cite{Merino05PRB}.  
Similar situation is realized in many molecular conductors as well, 
generally described by 2D anisotropic triangular lattice models at quarter-filling~\cite{Kino96JPSJ,Seo00JPSJ,Seo04CR}.
Such interplay between CO instability and geometrical frustration has also been 
discussed in triangular lattice systems such as LuFe$_2$O$_4$~\cite{Ikeda05Nature} 
and Na$_x$CoO$_2$~\cite{Motrunich04PRB}, 
spinel compounds such as AlV$_2$O$_4$~\cite{Matsuno01JPSJ} 
and LiV$_2$O$_4$~\cite{Fulde04JPCM}, and so on.

In these compounds, 
the CO transition, when it is realized, takes place accompanied with a large modulation 
in the lattice degree of freedom. 
For example, a first order structural phase transition 
is observed in $\theta$-ET$_2$RbZn(SCN)$_4$, 
where the ET molecules show displacements from their positions 
at high temperatures.~\cite{HMori98PRB,Miyagawa00PRB,Watanabe04JPSJ}
Ordering of Na$^+$ cations is suggested to be closely related to the CO state 
in Na$_{0.5}$CoO$_2$~\cite{Foo04PRL,Zandbergen04PRB}. 
In AlV$_2$O$_4$, a peculiar type of CO 
with a 3:1 ratio of V$^{2.5+\delta}$ and V$^{2.5-3\delta}$ ions 
has been suggested from a large rhombohedral distortion~\cite{Matsuno01JPSJ}. 
These instabilities can be considered as ``efforts" of 
the system to relax the geometrical frustration 
and to settle down in stable ordered states. 

On the other hand, when the compounds avoid such structural instabilities,  
conducting states are realized. 
PrBa$_2$Cu$_4$O$_8$ is metallic down to the lowest temperature~\cite{Horii00PRB}
but large charge fluctuation is seen in the NQR relaxations~\cite{Fujiyama03PRL}. 
In $\theta$-ET$_2$CsZn(SCN)$_4$ and in the rapid-cooled $\theta$-ET$_2$RbZn(SCN)$_4$, 
the electrical conductivity shows rather little temperature dependence 
down to about 20 K, below which a gradual increase of resitivity is observed~\cite{HMori98PRB}. 
There, an NMR measurement~\cite{Kanoda05JP4} suggests 
the existence of a ``charge glass"-like state 
where the dynamical CO fluctuation becomes frozen at low temperatures 
but not in a long-range way. 
Moreover, non-linear conductivity and insulator-like dielectric behavior are observed 
even when the systems are in the conductive state~\cite{Inagaki04JPSJ}. 

In this paper, we theoretically investigate 
such CO systems under strong geometrical frustration, 
by considering models which have been discussed previously to describe some of the materials above, 
i.e., 
geometrically frustrated quarter-filled extended Hubbard models 
which include  
competing intersite Coulomb interaction terms $V_{ij}$ in addition to the on-site $U$-term. 
Previous numerical works~\cite{Seo01PRB,Merino05PRB,Nishimoto03PRB,Watanabe06JPSJ} 
have revealed that CO insulating states in the unfrustrated region 
are melted when the frustration becomes strong, resulting in a metal 
even when the interaction strengths, $U$ and $V_{ij}$, are large. 
In this metallic state with large but frustrated $V_{ij}$, 
qualitative difference has hardly been found, mainly in static properties of the models, 
in comparison to a metal at ``weak-coupling" regime with small $V_{ij}$ (but with large $U$). 
We will see in the present paper that differences exist in dynamical quantities. 
Charge fluctuations of different CO patterns exist at finite energy in this metallic region, 
which is a novel property of CO system under strong geometrical frustration. 

The organization of this paper is as follows. 
In $\S$~\ref{sec_model}, 
1D and 2D frustrated extended Hubbard models which we consider, 
a summary of previous works on them, 
and the numerical Lanczos exact diagonalization method we use in this paper, are introduced. 
The results on the 1D and the 2D cases are shown in $\S$~\ref{sec_1D} and  \ref{sec_2D}, 
respectively, in which mainly the dynamical correlation functions are discussed. 
Section~\ref{sec_disc} is devoted to discussions and $\S$~\ref{sec_sum} to a summary.

\section{Model and Method}\label{sec_model}

\begin{figure}
\vspace*{1em}
\centerline{\includegraphics[width=6.5truecm]{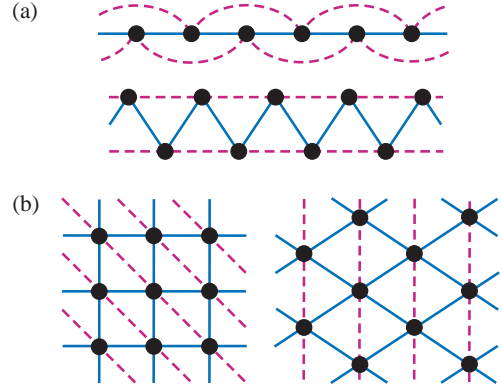}}
\caption{(Color online) (a) One-dimensional chain with next-nearest neighbor interaction (up) 
and zigzag ladder  (down), 
(b)  square lattice with next-nearest neighbor interaction along one diagonal (left) 
and anisotropic triangular lattice (right). 
These models, respectively, are equivalent with each other as seen in the figure. 
}
\label{fig1}
\end{figure}

We consider quarter-filled extended Hubbard models on two different lattice structures. 
The first is a 1D chain with nearest neighbor and next-nearest neighbor interactions, 
described as, 
\begin{align}
{\cal H}_{\rm 1D} &= \sum_{i\sigma} 
\{ - t_1(c^\dagger_{i, \sigma} c_{i+1,\sigma} + h.c.) 
 - t_2 (c^\dagger_{i,\sigma} c_{i+2,\sigma} + h.c.) \} \nonumber\\
 & \hspace{0.3cm} + \sum_{i} \left(U n_{i,\uparrow}  n_{i,\downarrow} 
+ V_1 n_i n_{i+1} + V_2 n_i n_{i+2} \right), 
\label{H1D}
\end{align}
where 
$\sigma$ is a spin index taking $\uparrow$ and $\downarrow$,
$c^{\dagger}_{i,\sigma}$ ($c_{i,\sigma}$) and $n_{i,\sigma}=c^\dagger_{i,\sigma} c_{i,\sigma}$ denote 
the creation (annihilation) operator and the number operator 
for the electron of spin $\sigma$ at the $i$-th site,
and $n_i=n_{i,\uparrow}+n_{i,\downarrow}$. 
The other is a 2D square lattice model with next-nearest neighbor interaction 
along one of the diagonals,  
\begin{align}
{\cal H}_{\rm 2D} = &\sum_{i\sigma} \{ 
- t_1 ( c^\dagger_{i,\sigma} c_{i+{\hat x},\sigma} 
+ c^\dagger_{i,\sigma} c_{i+{\hat y},\sigma} + h.c. ) \nonumber\\
& \hspace{3.5cm} - t_2 ( c^\dagger_{i,\sigma} c_{i+{\hat x}+{\hat y}, \sigma} + h.c. ) \} \nonumber\\
+ \sum_{i} &\left\{ U n_{i,\uparrow}  n_{i,\downarrow} 
+ V_1 \left( n_i n_{i+{\hat x}} + n_i n_{i+{\hat y}} \right) 
+ V_2 n_i n_{i+ {\hat x}+{\hat y}} \right\}, 
\label{H2D}
\end{align}
where ${\hat x}$ and ${\hat y}$ are the unit vectors with one lattice spacing 
along $x$ and $y$ directions, respectively. 
As shown in Fig.~\ref{fig1}, 
these models are equivalent to a 1D double-chain model with zigzag couplings 
between the chains, i.e., the zigzag ladder,  
and a 2D model on an anisotropic triangular lattice, respectively. 
The lattice constants in both models are set to unity in the following. 

The former 1D model at quarter-filling has been studied as an effective model for two different systems. 
One is for quarter-filled molecular conductors TMTSF$_2X$, 
where effects of long-range Coulomb interaction have been investigated~\cite{Kobayashi98JPSJ,Yoshioka01JPSJ}. 
In these studies $t_2$ is set to 0 while $V_2 < V_1$ is appropriate to the actual compounds 
since the Coulomb interactions are taken along the TMTSF chain. 
The other is for the Cu-O subunit of PrBa$_2$Cu$_4$O$_8$, 
mentioned in $\S$~\ref{sec_intro}, 
where the Cu sites are formed in a zigzag ladder structure~\cite{Seo01PRB,Nishimoto03PRB}. 
In this case $|t_1|\ll |t_2|$ 
due to the nearly 180$^\circ$ ($t_1$) and 90$^\circ$ ($t_2$) Cu-O-Cu angles,  
and $V_1 \simeq \sqrt{2} V_2$ suggested by the Cu-Cu distances. 
For $t_1=0$ or $t_2=0$, this 1D model at $U=\infty$ is mapped onto XXZ models~\cite{Seo01PRB} 
therefore the analogy between CO systems 
and localized spin systems under frustration 
metioned in $\S$~\ref{sec_intro} is straightforward here. 

The latter 2D model, on the other hand, together with its variants with lower symmetry, 
has been studied intensively 
motivated by recent experimental findings of CO in 2D molecular 
conductors~\cite{Merino05PRB,Seo00JPSJ,Seo04CR,TMori03JPSJ,Kaneko06JPSJ,Watanabe06JPSJ}. 
Among them, compounds with the highest symmetry, $\theta$-ET$_2X$, 
can be described by the model in eq.~(\ref{H2D}). 
Here, $X$ takes different monovalent anions, 
e.g., $MM'$(SCN)$_4$ with various metal elements in $MM'$ such as $M$=Rb, Cs, Tl 
and $M'$=Zn, Co~\cite{HMori98PRB}. 
In these $\theta$-ET$_2MM'$(SCN)$_4$ salts, 
the anisotropic shape of the HOMO forming the valence band at the Fermi level 
results in $t_1 \gg t_2$ while relative distances between molecules provide $V_1 \simeq V_2$. 

\begin{figure}
\vspace*{0.5em}
\centerline{\includegraphics[width=5truecm]{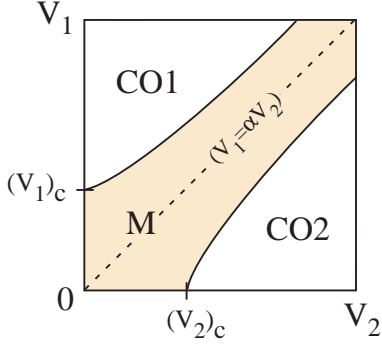}}
\caption{(Color online)
Schematic ground state phase diagram of frustrated extended Hubbard models, 
for fixed on-site Coulomb energy $U$ at a large value compared to the kinetic energy $t_{ij}$. 
CO1 and CO2 denote the charge ordered insulating phases favored by the competing intersite Coulomb repulsions 
$V_1$ and $V_2$, respectively, and M is the metallic phase. 
$(V_1)_{\rm c}$ and $(V_2)_{\rm c}$ are their critical values for the non-frustrated cases. 
Note that the detailed nature of the phases and their boundaries can be different 
depending on the lattice structures; 
for example the charge ordering and the metal-insulator transitions 
take place at different parameters in the 2D models~\cite{Calandra02PRB,Watanabe06JPSJ}.  
}
\label{fig2}
\end{figure}

Numerical studies~\cite{Seo01PRB,Nishimoto03PRB,Merino05PRB,Watanabe06JPSJ} 
on these 1D and 2D models have found 
a wide region of metallic phase between two CO insulating phases 
in their ground state phase diagrams, as shown in Fig.~\ref{fig2}. 
This is drawn by varying $V_1$ and $V_2$ ( $\lesssim U/2$) 
while $U$ is fixed at a large value compared to the kinetic term $t_{ij}$. 
In the two CO phases, Wigner crystal-type CO states are stabilized 
due to either $V_1$ or $V_2$: charge localization occurs in order to avoid 
the Coulomb repulsion along the nearest neighbor or the next-nearest neighbor bonds. 
In the 1D case, they are the ``[1010]" and ``[1100]" 
(or $4\kf$ and $2\kf$ CO states in terms of $\kf=\pi/4$ for $t_2=0$) patterns, 
while in the 2D case, the checkerboard-type and the stripe-type patterns, respectively. 
The metallic phase is realized in the $V_1 \simeq \alpha V_2$ region, 
where $\alpha=1$ ($\alpha=2$) for the 1D (2D) model, 
due to strong geometrical frustration in $V_{ij}$. 
The line along $V_1 = \alpha V_2$ is the ``fully" frustrated line 
where the corresponding Ising models $\sum_{\langle ij \rangle} V_{ij} \sigma_i \sigma_j$ 
have a spin disordered ground state due to macroscopic degeneracy. 

In this work, in order to incorporate quantum fluctuation which is crucial 
in describing this frustration-induced metallic state, 
we use standard numerical Lanczos exact diagonalization method for finite size clusters~\cite{Dagotto94RMP}. 
The electron number $n$ is fixed at quarter-filling, namely, $n=L/2$ 
where $L$ is the total number of sites in the cluster.  
We mainly consider the dynamical charge/spin correlation functions, 
\begin{align}
N({\boldsymbol q},\omega) &=\sum_\nu |\langle \nu|N_{\boldsymbol q} |0\rangle|^2 \delta(\omega-E_\nu+E_0) \\
S({\boldsymbol q},\omega) &= \sum_\nu |\langle \nu|S^z_{\boldsymbol q} |0\rangle|^2 \delta(\omega-E_\nu+E_0)
\end{align} 
where $E_0$ and $E_\nu$ are the ground state ($|0\rangle$) and excited state ($|\nu\rangle$)
energies of the system, respectively. 
$N_{\boldsymbol q}$ and $S^z_{\boldsymbol q}$ are Fourier transformations 
of the electron density with respect to the mean value ${\bar n}=n/L=1/2$, $N_i=n_i - {\bar n}$, and 
the $z$-compenent of the spin operator, 
$S^z_i = (c_{i\uparrow}^\dagger c_{i\uparrow} - c^\dagger_{i\downarrow} c_{i\downarrow} )/2$, 
i.e., 
$N_{\boldsymbol q}= {L^{-1/2}} \sum_i e^{\rm{i}{\boldsymbol q}\cdot{\boldsymbol R}_i} N_i$ and
$S^z_{\boldsymbol q}={L^{-1/2}} \sum_i e^{\rm{i}{\boldsymbol q}\cdot{\boldsymbol R}_i} S^z_i $. 

In order to consider the CO patterns expected from the previous works, 
including the possible spin patterns, 
$L=4m$ ($m$: integer) site chains are required for the 1D case, while in the 2D case 
the $L=4\times4=16$ site square-shaped 
cluster is the only available size when considering the matching of the 
cluster shape with the periodicities of such CO/spin order, and the computational limitation; 
for the 1D case the $L=16$ site chain is used. 
The antiperiodic boundary condition is choosen for both cases  
to give the closed shell in their non-interacting bands~\cite{NoteLevel}. 
Note that the latter 2D cluster rules out long-period charge modulations 
recently found in the frustrated metallic state 
in this 2D model~\cite{TMori03JPSJ,Kaneko06JPSJ,Watanabe06JPSJ}. 

In the following calculations we set $t_1$ as unity, 
and fix $t_2=0$ and $U=10$ 
while varying the intersite Coulomb energies $V_1$ and $V_2$.
Although some of the properties of the models might be modified 
by different values of $t_2$, 
as far as the properties discussed in this paper 
we expect that they will not distract our main conclusions. 

\section{One-dimensional case}\label{sec_1D}

\begin{figure}[b]
 \centerline{\includegraphics[width=7truecm]{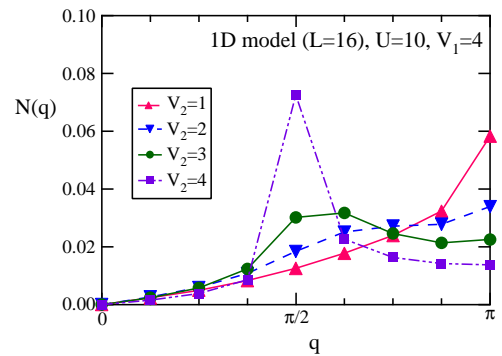}}
\vspace*{-1em}
\caption{(Color online)
Profile of $N({\boldsymbol q}$) in the one-dimensional frustrated extended Hubbard model ($L=16$) 
at quarter-filling, for fixed $U=10$, $V_1=4$ and for different values of $V_2$. 
} 
\label{fig3}
\end{figure}

Before discussing the dynamical properties of the 1D model, 
let us remark a result on static properties that was not noticed in the previous studies: 
an incommensurate (IC) peak in 
the equal-time charge correlation function, 
$N({\boldsymbol q})=  {L^{-1}} \langle 0| N_{- {\boldsymbol q}} N_{\boldsymbol q} |0 \rangle 
= {L^{-1}} \sum_{ij} e^{\rm{i} {\boldsymbol q} \cdot {\boldsymbol R}_{ij} }\langle 0|
N_i  N_j |0 \rangle$, where ${\boldsymbol R}_{ij}$ denotes 
the vector connecting ${\boldsymbol R}_i$ and ${\boldsymbol R}_j$, 
when the system is in the frustration-induced metallic phase. 
In Fig.~\ref{fig3}, 
profiles of $N({\boldsymbol q})$ for fixed $V_1=4$ are plotted, 
for different values of $V_2$. 
For this value of $V_1$, by increasing the value of $V_2$ from 0, 
the ground state changes from the [1010]-type $4\kf$ CO insulator to 
the metal in the intermediate region, 
and then to the [1100]-type $2\kf$ CO insulator~\cite{Nishimoto03PRB}. 
These two CO states are characterized by commensurate $N({\boldsymbol q})$ peaks
at ${\boldsymbol q}=\pi$ and $\pi/2$, 
typically seen in Fig.~\ref{fig3} for $V_2=1$ and $V_2=4$, respectively. 
However, for $V_2=2$ one can see a double peak structure 
at ${\boldsymbol q}=\pi$ and at an IC value ${\boldsymbol q}\simeq 3\pi/4$
(which may be a convolution of ${\boldsymbol q}=\pi$ and ${\boldsymbol q} \simeq 5\pi/8$ peaks).  
For $V_2=3$ the IC peak shifts to ${\boldsymbol q}\simeq 5\pi/8$  
and dominates over the ${\boldsymbol q}=\pi$ peak. 

We note that this result does not immediately point to the existence 
of an IC charge density modulation in the ground state of this model. 
Although we find such IC peaks also in different 
cluster sizes, our study is unsufficient 
for investigating the long distance behavior of this IC 
correlation, since considerably large system sizes are needed 
to make many ${\boldsymbol q}$ values available and to proceed finite size scaling of them. 
Considering the density-matrix-renormalization-group 
results by Nishimoto \et~\cite{Nishimoto03PRB} supporting the Tomonaga-Luttinger liquid (TLL) properties 
in this metallic regime, 
we expect that this IC correlation would decrease exponentially 
and vanish at long distances since the TLL implies that 
low-lying excitations are described by modes at ${\boldsymbol q}=m\kf$ ($m$: integer). 
Still, this result is notable since similar long period modulations 
are found by different authors in the 2D models 
in the frustrated region~\cite{TMori03JPSJ,Kaneko06JPSJ,Watanabe06JPSJ}, 
although we cannot find them in the 2D cluster that we study 
as pointed out in $\S$~\ref{sec_model}. 
We will mention about it in $\S$~\ref{sec_disc}.

\begin{figure}
\vspace*{1em}
 \centerline{\includegraphics[width=7.3truecm]{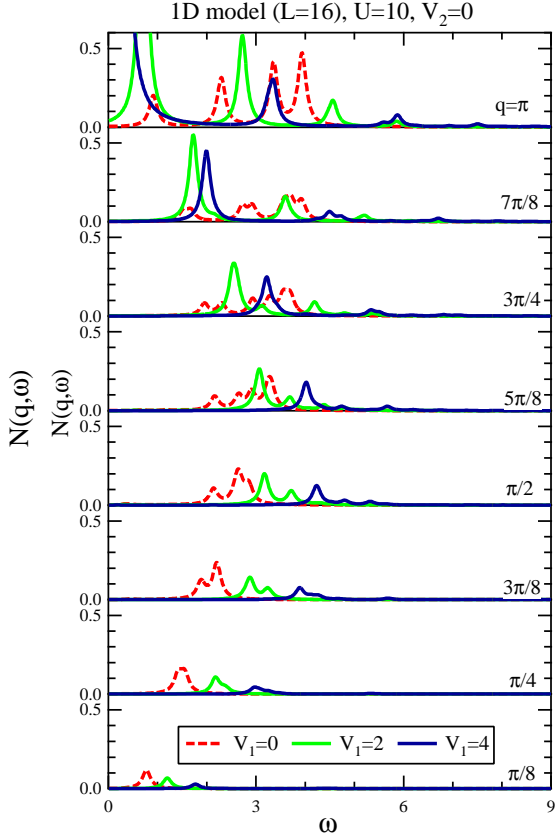}}
\vspace*{-1em}
\caption{(Color online)
$N({\boldsymbol q},\omega)$ 
for the quarter-filled one-dimensional extended Hubbard model ($L=16$)
without frustration ($V_2=0$) at fixed $U=10$, for different values of $V_1$.  
The system size is $L=16$ and the antiperiodic boundary condition is used. 
The Lorentzian broadening of the delta functions is $\eta=0.1$. 
} 
\label{fig4}
\end{figure}

\begin{figure}
\vspace*{1em}
 \centerline{\includegraphics[width=7.3truecm]{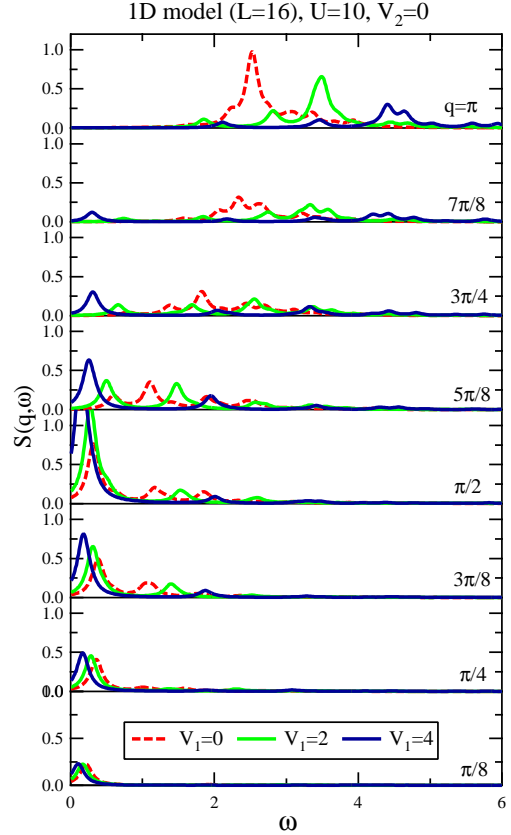}}
\vspace*{-1em}
\caption{(Color online)
$S({\boldsymbol q},\omega)$
for the quarter-filled one-dimensional extended Hubbard model ($L=16$)
without frustration  ($V_2=0$) at fixed $U=10$, for different values of $V_1$.  
The Lorentzian broadening of the delta functions is $\eta=0.1$. 
} 
\label{fig5}
\end{figure}

Next, we proceed to the dynamical properties of this model. 
For clarity, we first show 
how the quantities behave in the case without frustration, 
and then switch on the frustration. 
In Figs.~\ref{fig4} and~\ref{fig5}, 
$N({\boldsymbol q},\omega)$ and $S({\boldsymbol q},\omega)$
at $V_2=0$, i.e., the usual 1D extended Hubbard model, for several values of $V_1$ are shown. 
For $V_1=V_2=0$, the model is just the 1D Hubbard model 
and the obtained spectra are 
consistent with the large $U$ study by Ogata and Shiba~\cite{Ogata90PRB}  
based on the Bethe ansatz, 
as well as those of the 1D $t$-$J$ model~\cite{Bares91PRB,TohyamaPRL95}, 
which can be considered as an effective model for the Hubbard model at large $U$. 
The spectra are composed of 
low-lying modes in the charge sector at $4\kf$ and in the spin sectors at $2\kf$~\cite{TohyamaPRL95}, 
i.e., at ${\boldsymbol q}=\pi$ and at ${\boldsymbol q}=\pi/2$ for quarter-filling, respectively.  
When $V_1$ is increased, 
in $N({\boldsymbol q},\omega)$ a strong softening of the ${\boldsymbol q}=\pi$ mode 
is seen, signaling the instability toward the $4\kf$ CO. 
At the same time, in $S({\boldsymbol q},\omega)$ the low-lying ${\boldsymbol q}=\pi/2$ mode 
becomes sharper and around ${\boldsymbol q}=\pi/2$ 
the dispersion is more symmetric with respect to this ${\boldsymbol q}$ value. 
One can interpret this behavior as a crossover to a more localized 
spin-like state due to the charge localization, 
with a well-defined spin-wave dispersion centered at ${\boldsymbol q}=\pi/2$. 
The difference in the energy scale for the charge and spin degrees of freedoms 
is noticeably seen in the dispersions of these modes as distinct total widths of them, 
which is due to the fact that the CO is determined by the Coulomb energies $U$ and $V$ 
while the spins are interacting with an effective Heisenberg coupling, 
$J$~\cite{Tanaka05JPSJ}. 

\begin{figure}
\vspace*{1em}
 \centerline{\includegraphics[width=7.5truecm]{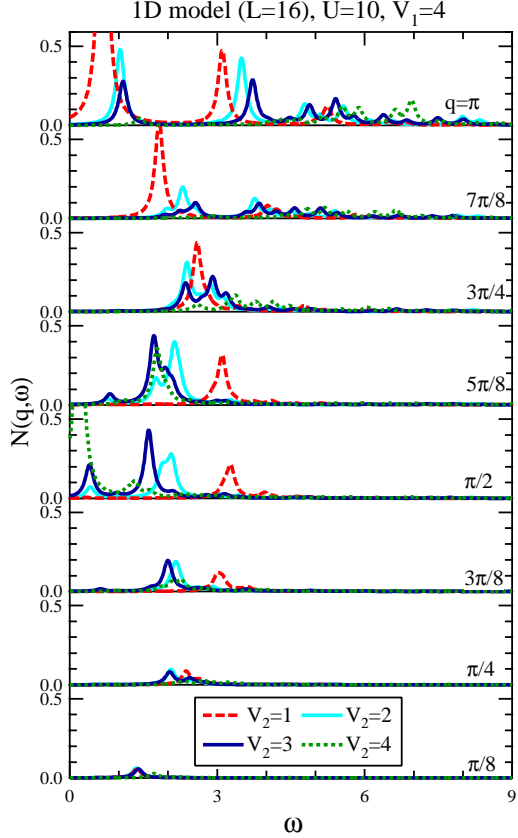}}
\vspace*{-1em}
\caption{(Color online)
$N({\boldsymbol q},\omega)$
for the quarter-filled one-dimensional extended Hubbard model ($L=16$)
with frustration at fixed $U=10$, $V_1=4$, and for different values of $V_2$.  
The Lorentzian broadening of the delta functions is $\eta=0.1$. 
} 
\label{fig6}
\end{figure}
\begin{figure}
\vspace*{1em}
 \centerline{\includegraphics[width=7.5truecm]{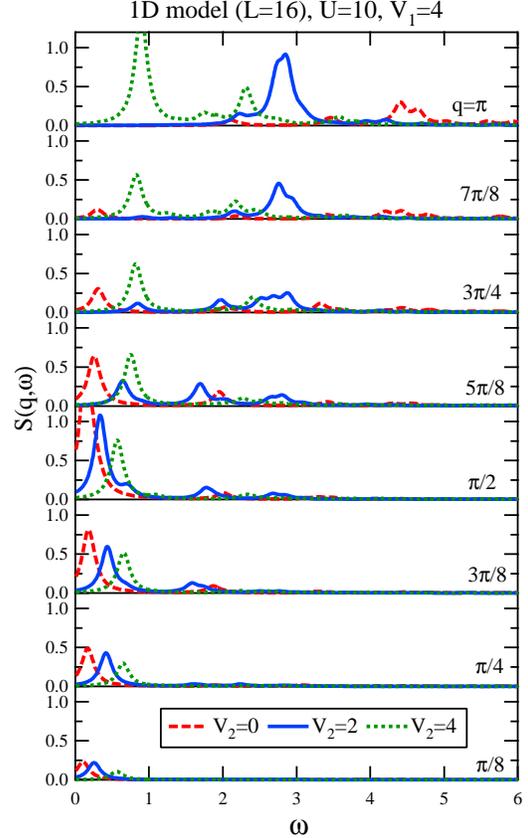}}
\vspace*{-1em}
\caption{(Color online)
$S({\boldsymbol q},\omega)$
for the quarter-filled one-dimensional extended Hubbard model ($L=16$)
with frustration at fixed $U=10$, $V_1=4$, and for different values of $V_2$.  
The Lorentzian broadening of the delta functions is $\eta=0.1$. 
} 
\label{fig7}
\end{figure}

The variation of these spectra 
by the inclusion of frustration controlled by the value of $V_2$ 
is shown in Figs.~\ref{fig6} 
and \ref{fig7}. 
While fixing at $V_1=4$, by increasing $V_2$ 
the charge fluctuation seen in $N({\boldsymbol q},\omega)$ 
shows a drastic modification. 
As seen in Fig.~\ref{fig6}, 
the spectrum for $V_2=4$, at which the system 
is expected to be in the region of the 2$\kf$ CO state, 
a low-lying mode is mostly concentrated at ${\boldsymbol q}=\pi/2$. 
This is similar to the case of the 4$\kf$ CO state that we have observed in Fig.~\ref{fig4} 
with the ${\boldsymbol q}=\pi$ mode. 
In between, in the frustration-induced metallic region 
for $V_2=2$ and for $V_2=3$, we can see that the spectrum smoothly crosses over 
between the dispersions characterstic of the two different CO patterns. 
Interestingly enough, 
prominent modes at different ${\boldsymbol q}$ values are simultaneously seen in this region. 
For example, for $V_2=2$, peaks at 
${\boldsymbol q}=\pi$ and ${\boldsymbol q}=5\pi/8$ 
are larger compared to the other neighboring ${\boldsymbol q}$ values. 
As for the spectrum for $V_2=3$, the ${\boldsymbol q}=\pi/2$ peak 
becomes comparable to the latter peak. 
These match with the IC peak in $N({\boldsymbol q})$ mentioned above, 
i.e., the integration of $N({\boldsymbol q},\omega)$ along the $\omega$-axis. 
These results suggest that fluctuations of different CO states 
co-exist as dynamical charge correlation in the frustration-induced metallic region. 
We note that these are consistent with the TLL state at low energy 
since the energy positions of the dispersion 
always show minima at either ${\boldsymbol q}=\pi/2$ or $\pi$, 
matching the above condition ${\boldsymbol q}=m\kf$ ($m$: integer),  
even when the peak heights are highest at a different ${\boldsymbol q}$. 

The spin degree of freedom is consistent with such features. 
As seen in Fig.~\ref{fig7}, 
the spectra with the spin-wave mode centered at ${\boldsymbol q}=\pi/2$ 
seen in the $V_2=0$ case for the $4\kf$ CO state 
show a smooth crossover to the spin exitation at ${\boldsymbol q}=\pi$ characteristic of 
the $2\kf$ CO state at large $V_2$\cite{Kobayashi98JPSJ,Yoshioka01JPSJ}. 
In the latter state the pronounced peak is seen at ${\boldsymbol q}=\pi$ 
although it stays at finite energy even when $V_2$ is largely increased, 
which is due to the expected spin gap, as the [1100] configuration of 
charge localization produces pairs of singlet formation~\cite{Nishimoto03PRB}. 
In the frustration-induced TLL region, 
the $S({\boldsymbol q},\omega)$ spectrum resembles 
to that in the unfrustrated case at $V_1=V_2=0$ in Fig.~\ref{fig5}, 
in contrast with their clear difference in $N({\boldsymbol q},\omega)$. 

\section{Two-dimensional case}\label{sec_2D}

\begin{figure}
\vspace*{1em}
\centerline{\includegraphics[width=7.5truecm]{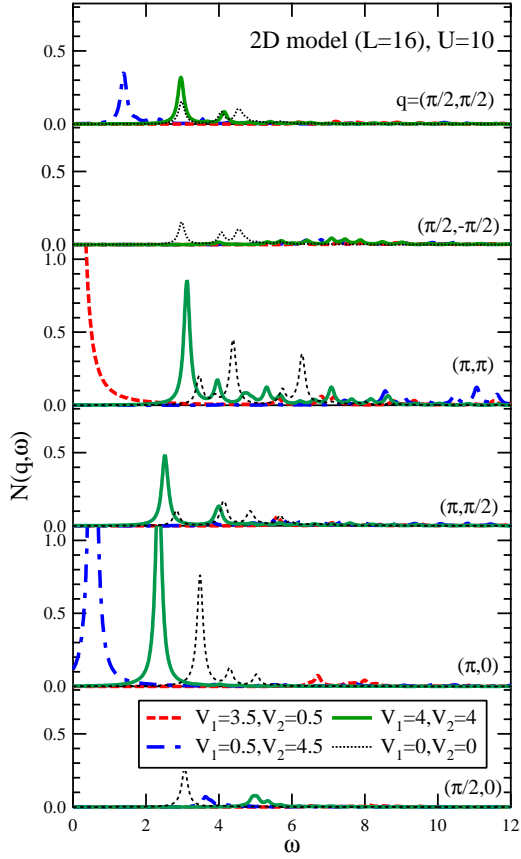}}
\vspace*{-1em}
\caption{(Color online)
$N({\boldsymbol q},\omega)$
for the quarter-filled two-dimensional extended Hubbard model ($L=16$)
with frustration at $U=10$, and for different sets of ($V_1$,$V_2$).  
The Lorentzian broadening of the delta functions is $\eta=0.1$. 
} 
\label{fig8}
\end{figure}

\begin{figure}
\vspace*{1em}
\centerline{\includegraphics[width=7.5truecm]{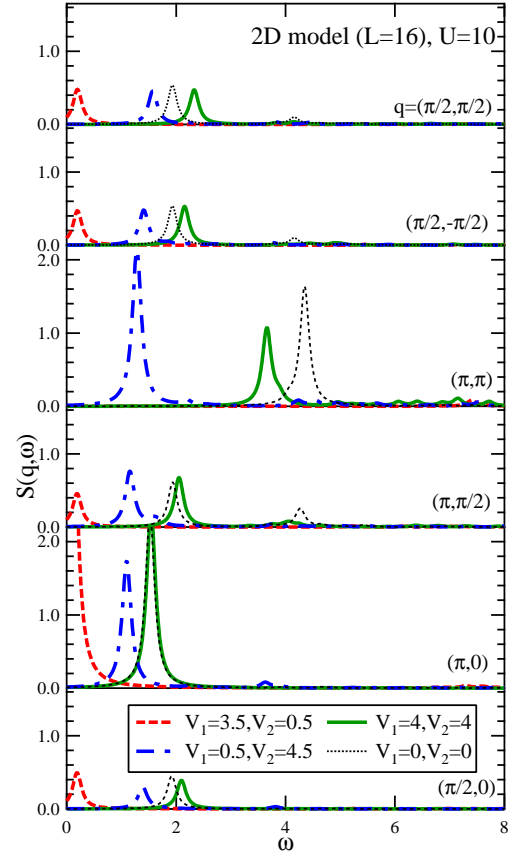}}
\vspace*{-1em}
\caption{(Color online)
$S({\boldsymbol q},\omega)$
for the quarter-filled two-dimensional extended Hubbard model ($L=16$)
with frustration at $U=10$, and for different sets of ($V_1$,$V_2$).  
The Lorentzian broadening of the delta functions is $\eta=0.1$. 
} 
\label{fig9}
\end{figure}

In the 2D model, 
the crossover behavior between two CO insulating states 
is also observed in the dynamical properties of the system. 
In Figs.~\ref{fig8} and \ref{fig9}, 
$N({\boldsymbol q},\omega)$ and $S({\boldsymbol q},\omega)$ spectra~\cite{NoteQ} are shown 
for different $(V_1,V_2)$ parameter sets across the phase diagram in Fig.~\ref{fig2} 
(see also Fig.~1 of Ref.~\ref{Merino05PRB}). 
When one starts from the checkerboard-type CO insulating state in the (large $V_1$, small $V_2$) region, 
by decreasing $V_1$ and increasing $V_2$, 
the ground state transforms to the intermediate metallic 
and then to the stripe-type CO insulating phase~\cite{Merino05PRB}. 
The two CO states are characterized by wave vectors 
$\boldsymbol{q}=(\pi,\pi)$ for the checkerboard pattern~\cite{Ohta94PRB,Calandra02PRB} 
and $\boldsymbol{q}=(\pi,0)$ for the stripe pattern\cite{Merino05PRB,Seo00JPSJ}. 
These are seen in the large peaks at $\omega\simeq 0$ 
in the $N({\boldsymbol q},\omega)$ spectra, 
as typically seen in Fig.~\ref{fig8} for $(V_1,V_2)=(3.5,0.5)$ and $(0.5,4.5)$, respectively. 
At other wave vectors, peaks are relatively small and located at higher energies. 

On the other hand, in the frustration-induced metallic phase 
at $(V_1,V_2)=(4,4)$, both of the two peaks 
at $\boldsymbol{q}=(\pi,\pi)$ and $(\pi,0)$ are prominent 
at the same time, but located at higher energies than those in the CO states. 
It shows that these modes are still present, 
although, 
due to the limited number of data for different wave vectors in contrast with the 1D case, 
how the curve of the dispersion would be in the thermodynamic limit is difficult to judge; 
actually the positions of the peaks among the available $\boldsymbol{q}$ values 
show little variation in energy.  
This is more clearly seen when we compare with 
the $N({\boldsymbol q},\omega)$ data for $(V_1,V_2)=(0,0)$, 
shown in Fig.~\ref{fig8} for reference, 
which is more dispersive and continuum-like peaks show up. 
The clear distinction between these two spectrum in the metallic region 
leads us to conclude that charge fluctuations for different CO states co-exist 
in the strongly frustrated region. 
This is similar to the 1D case where different CO modes 
were observed in the intermediate metallic phase, 
but with more clear dispersions (Fig.\ref{fig6}).

The $S({\boldsymbol q},\omega)$ spectra, again, are consistent with such crossover behavior, 
although its behavior is apparently rather complicated.  
In the checkerboard CO state the spins are located on the charge rich sites
and antiferromagnetically coupled. 
Due to the cluster shape we take, the spin pattern can be 
described by different $\boldsymbol{q}$ values such as $\boldsymbol{q}=(\pi/2,\pi/2)$ or $(\pi,0)$. 
The latter is developped as a pronounced low-energy peak in Fig.~\ref{fig9} 
for $(V_1,V_2)=(3.5,0.5)$. In our calculation it is found that 
spin pattern with either of the above $\boldsymbol{q}$ is developped  (or both) 
can depend on the parameters of the model and boundary conditions. 
This is probably a finite size effect, but we could not specify the actual origin of it. 
On the other hand, in the stripe CO state 
the 1D chains are coupled antiferromagnetically along the chains. 
The interchain configuration, this time, is in fact expected to depend on the parameters of the model 
in the thermodynamic limit, 
suggested by previous mean-field calculations~\cite{Seo00JPSJ,Kaneko06JPSJ}. 
The peaks seen in both $\boldsymbol{q}=(\pi,0)$ and $(\pi,\pi)$ in the $(V_1,V_2)=(0.5,4.5)$ 
indicate that the antiferromagnetic stripes have ferromagnetic correlation in their transverse 
directions. However, again, 
we find that not only the parameters of the model but also the boundary condition 
can change this transverse phase, therefore the results about the actual 
$\boldsymbol{q}$ value in this small cluster size should not be taken conclusively. 
Note the difference in the energy scale between
$N({\boldsymbol q},\omega)$ and $S({\boldsymbol q},\omega)$ spectra 
in these CO states, 
where small but noticeable peaks exist at higher energy in the former 
while almost all spectral weight is concentrated at $\omega < 2$ in the latter; 
this is analogous to the 1D case. 
In between, in the intermediate metallic region 
the spectrum is composed of modes 
surviving at the above $\boldsymbol{q}$ values, 
as seen in the $(V_1,V_2)=(4,4)$ data. 
The $S({\boldsymbol q},\omega)$ spectrum at $(V_1,V_2)=(0,0)$ 
is slightly more dispersive and extended to higher energy region 
than the rather flat dispersion in this $(V_1,V_2)=(4,4)$ data 
for the strongly frustrated metal. 

\section{Discussions}\label{sec_disc}

The IC correlation we found in the 1D model in $\S$~\ref{sec_1D} 
is a typical consequence of geometrical frustration in general. 
We suspect that it would also appear in 2D model as well 
if one could have larger system sizes available. 
Such a state supposedly correspond to the so-called ``three-fold" state found 
in mean-field studies~\cite{TMori03JPSJ,Kaneko06JPSJ} 
as well as in a variational Monte-Carlo study~\cite{Watanabe06JPSJ} of this 2D model. 
A recent RPA study~\cite{KurokiPrivate} found that the position of $\boldsymbol{q}$ 
where the charge susceptibility shows a maximum is shifted 
by changing the range of intersite Coulomb interaction terms and their values, 
and can be adjusted to that found in X-ray experiments on $\theta$-ET$_2X$ 
as diffusive rods, i.e., as a short range order. 
Such an IC correlation is found in spin systems as well. 
Exact diagonalization calculations for finite size 1D clusters of the so-called Majumdar-Ghosh model, 
i.e., the $S=1/2$ Heisenberg model with next-nearest neighbor coupling, 
show such IC peak in its spin structure factor $S(\boldsymbol{q})$~\cite{Tonegawa87JPSJ}, 
and an IC spin ordered ground state is actually found experimentally 
in a quasi-1D $S=1/2$ spin system Cs$_2$CuCl$_4$~\cite{Coldeay96JPCM}. 
It is interesting to search for such an IC state attaining long range order of 
charge modulation in the CO materials as well. 
This will bring about different aspects from the spin systems, 
since for the CO systems the interplay with the metal-insulator transition 
would be an interesting issue. 

We consider that the co-existence of different CO modes 
in the dynamical spectra for the strongly frustrated metallic region 
that we have shown in both 1D and 2D models,  
is a general characteristic of geometrical frustrated CO systems. 
In the 2D model, a variational Monte Carlo simulation 
by Watanabe and Ogata~\cite{Watanabe06JPSJ} suggested that 
variational states with different CO patterns, 
i.e., the checkerboard, the stripe, and the threefold-type, 
have very close ground state energies in the frustrated metallic region. 
Although their data is for static ground states while ours is for dynamic quantities, 
we consider that we are detecting different aspects of the same phenomena. 

The flatness of the dispersions in this region for the 2D model shown in $\S$~\ref{sec_2D}, 
both in $N({\boldsymbol q},\omega)$ and $S({\boldsymbol q},\omega)$,  
is a peculiar point. In the 1D case, in contrast, 
the width of the dispersion in $N({\boldsymbol q},\omega)$ is not much reduced, 
and that in $S({\boldsymbol q},\omega)$ hardly differs, 
compared with the ``weak-coupling" metal at $V_1=V_2=0$. 
This indicates that the charge and spin excitations are very localized 
in real space for the 2D system, while it can propagate to some extent in the 1D case. 
It is due to the dimensionality effect, 
since, e.g., an addition of charge, or a flipping of spin in the 2D model 
would distract the environment due to the strong coupling strenghs, 
and therefore cannot propagate into the ``bulk". 
The flatness results in a ``downward shift of spectral weight" 
compared with the $V_1=V_2=0$ data, 
which is pointed out by Ramirez~\cite{FrustrationReview} 
to be a tendance of geometrically frustrated spin systems. 

The relation between our results with the experimental data 
showing the existence of charge fluctuation at very low energy scale 
detected by NQR/NMR is still obscure. 
By NQR measurements on PrBa$_2$Cu$_4$O$_8$, 
Fujiyama \et \cite{Fujiyama03PRL} have evaluated the fraction of such fluctuating (or frozen) CO to be very small 
compared to the bulk, and they have deduced this to be related with small amount of impurities. 
Kanoda \et \cite{Kanoda05JP4} have proposed the existence of a charge glass state in $\theta$-ET$_2X$, 
an analogy of the spin glass state in spin systems, therefore again 
points to the role of impurities under the presence of geometrical frustration. 
In addition, CO is in general coupled with the lattice, not included in our model, 
which may also be important in understanding such phenomena. 
However, the behavior of the 1D and 2D frustrated extended Hubbard models 
we discussed in this paper would be helpful in analyzing such experiments to start with. 

\section{Summary}\label{sec_sum}

In summary, 
we have theoretically investigated CO systems under strong geometrical frustration, 
by treating frustrated extended Hubbard models on 1D and 2D lattice structures. 
Charge fluctuations of different types of CO states are found to 
co-exist in the frustration-induced metallic region of these models, 
showing up as dynamical modes in the correlation function spectra. 
We have also found incommensurate charge correlation in the 1D model, 
and deduced that it is a general tendence in other frustrated CO systems as well. 

\section*{Acknowledgment}
We thank H. Fukuyama, K. Kanoda, H. Watanabe, and K. Yakushi 
for valuable discussions. 
This work is supported by 
the Grant-in-Aid for Scientific Research on Priority Area of
``Molecular Conductors" from MEXT, 
and J. M. acknowledges support from the Ram{\'o}n y Cajal program in Spain and MEC 
under contract: CTQ2005-09385-c03-03.


\begin{thebibliography}{99} 
%
\bibitem{Verwey39Nature} E. J. W. Verwey: \jo{Nature (London)}{144}{1939}{327}.
%
\bibitem{Fe3O4Review}For a review, 
N. Tsuda, K. Nasu, A. Fujimori, and K. Shiratori: 
{\it Electronic Conduction in Oxides}, 2nd ed. (Springer-Verlag, Berlin, 2000), p.~243.
%
\bibitem{Novak00PRB}
P. Novak, H. Stepankova, J. Englich, J. Kohout, and V. A. M. Brabers: \jo{\PRB}{61}{2000}{1256}. 
%
\bibitem{Garcia00PRB}
J. Garc{\' \i}a, G. Sub{\' \i}as, M. G. Proietti, J. Blasco, H. Renevier, J. L. Hodeau, and Y. Joly: 
\jo{\PRB}{63}{2000}{054110}. 
%
\bibitem{Todo01JAP}
S. Todo, N. Takeshita, T. Kanehara, T. Mori, and N. M{$\hat{\rm o}$}ri: \jo{J. Appl. Phys.}{89}{2001}{7347}.
%
\bibitem{Seo02PRB} H. Seo, M. Ogata, and H. Fukuyama: \jo{\PRB}{65}{2002}{085107}. 
%
\bibitem{Anderson56PR} P. W. Anderson: \jo{\PR}{102}{1956}{1008}.
%
\bibitem{FrustrationReview}
A. P. Ramirez: in \textit{Handbook of Magnetic Materials}, edited by K. H. J. Buschow (Elsevier, Amsterdam, 2001), 
Vol. 13, p. 423. 
%
\bibitem{Horii00PRB}
S. Horii, U. Mizutani, H. Ikuta, Y. Yamada, J. H. Ye, A. Matsushita, N. E. Hussey, H. Takagi, and I. Hirabayashi:
\jo{\PRB}{61}{2000}{6327}. 
%
\bibitem{Seo01PRB} H. Seo and M. Ogata: \jo{\PRB}{64}{2001}{113103}; \jo{\PRB}{65}{2002}{169902(E)}. 
%
\bibitem{HMori98PRB} H. Mori, S. Tanaka, and T. Mori: \jo{\PRB}{57}{1998}{12023}. 
%
\bibitem{Merino05PRB}\label{Merino05PRB} J. Merino, H. Seo, and M. Ogata: \jo{\PRB}{71}{2005}{125111}. 
%
\bibitem{Kino96JPSJ} H. Kino and H. Fukuyama: \jo{\JPSJ}{65}{2158}{1996}. 
%
\bibitem{Seo00JPSJ} H. Seo: J. Phys. Soc. Jpn. {\bf 69}, 805 (2000).
%
\bibitem{Seo04CR} H. Seo, C. Hotta, and H. Fukuyama: \jo{\CR}{104}{2004}{5005}. 
%
\bibitem{Ikeda05Nature}
N. Ikeda, H. Ohsumi, K. Ohwada, K. Ishii, T. Inami, K. Kakurai, Y. Murakami, K. Yoshii, S. Mori, Y. Horibe, 
and H. Kit{\^ o}: 
\jo{Nature}{436}{2005}{1136}. 
%
\bibitem{Motrunich04PRB} O. I. Motrunich and P. A. Lee: \jo{\PRB}{69}{2004}{214516}. 
%
\bibitem{Matsuno01JPSJ}
K. Matsuno, T. Katsufuji, S. Mori, Y. Moritomo, A. Machida, E. Nishibori, M. Takata, M. Sakata, N. Yamamoto, 
and H. Takagi: 
\jo{\JPSJ}{70}{2001}{1456}. 
%
\bibitem{Fulde04JPCM}
P. Fulde: J. Phys. Condens. Matter {\bf 16} (2004) S591.
%
\bibitem{Miyagawa00PRB} K. Miyagawa, A. Kawamoto, and K. Kanoda: \jo{\PRB}{62}{2000}{7679}.  
%
\bibitem{Watanabe04JPSJ}
M. Watanabe, Y. Noda, Y. Nogami, and H. Mori: \jo{\JPSJ}{73}{2004}{116}. 
%
\bibitem{Foo04PRL} 
M. L. Foo, Y. Wang, S. Watauchi, H. W. Zandbergen, T. He, R. J. Cava, and N. P. Ong: 
\jo{\PRL}{92}{2004}{247001}.
%
\bibitem{Zandbergen04PRB}
 H. W. Zandbergen, M. L. Foo, Q. Xu, V. Kumar, and R. J. Cava: \jo{\PRB}{70}{2004}{024101}. 
%
\bibitem{Fujiyama03PRL} S. Fujiyama, M. Takigawa, and S. Horii: \jo{\PRL}{90}{2003}{147004}. 
%
\bibitem{Kanoda05JP4}
K. Kanoda, K. Ohnou, M. Kodama, K. Miyagawa, T. Itou,  and K. Hiraki: J. de Physique IV {\bf 131} (2005) 21.
%
\bibitem{Inagaki04JPSJ} 
K. Inagaki, I. Terasaki, H. Mori, and T. Mori: \jo{\JPSJ}{73}{2004}{3364}.
%
%
\bibitem{Nishimoto03PRB}
 S. Nishimoto and Y. Ohta: \jo{\PRB}{68}{2003}{235114}; 
 S. Ejima, F. Gebhard, S. Nishimoto, and Y. Ohta: \jo{\PRB}{72}{2005}{033101}. 
%
\bibitem{Watanabe06JPSJ}\label{Watanabe06JPSJ} H. Watanabe and M. Ogata: \jo{\JPSJ}{75}{2006}{063702}.
%
\bibitem{Kobayashi98JPSJ} N. Kobayashi, M. Ogata, and K. Yonemitsu: \jo{\JPSJ}{67}{1998}{1098}.
%
\bibitem{Yoshioka01JPSJ} H. Yoshioka, M. Tsuchiizu, and Y. Suzumura: \jo{\JPSJ}{70}{2001}{762}.
%
\bibitem{TMori03JPSJ}\label{TMori03JPSJ} T. Mori: \jo{\JPSJ}{72}{2003}{1469}.
%
\bibitem{Kaneko06JPSJ}\label{Kaneko06JPSJ} M. Kaneko and M. Ogata:\jo{\JPSJ}{75}{2006}{014710}.  
%
\bibitem{Calandra02PRB}\label{Calandra02PRB}
M. Calandra, J. Merino, and R. H. McKenzie: Phys. Rev. B {\bf 66} (2002) 195102. 
%
\bibitem{Dagotto94RMP} 
For example, see E. Dagotto:
Rev. Mod. Phys. {\bf 66} (1994) 763, and references therein. 
%
\bibitem{NoteLevel}For the antiperiodic boundary condition, 
we find no level crossing in both the 1D and the 2D models 
for the parameter sets we used in this paper. 
%
\bibitem{Ogata90PRB} M. Ogata and H. Shiba: \jo{\PRB}{41}{1990}{2326}. 
%
\bibitem{Bares91PRB} P.-A. Bares, G. Blatter, and M. Ogata: \jo{\PRB}{44}{1991}{130}.
%
\bibitem{TohyamaPRL95} T. Tohyama, P. Horsch, and S. Maekawa: \jo{\PRL}{74}{1995}{980}.
%
\bibitem{Tanaka05JPSJ} Y.\ Tanaka and M.\ Ogata: \jo{\JPSJ}{74}{2005}{3283}.
%
\bibitem{NoteQ} These quantities at $\boldsymbol{q}=(\pi/2,\pi/2)$ and 
at $\boldsymbol{q}=(\pi/2,-\pi/2)$ become different as $V_2$ is introduced, 
in contrast to the square lattice case. 
Note that those at $\boldsymbol{q}=(\pm\pi,\pm\pi/2)$ and at $\boldsymbol{q}=(\pm\pi/2,\pm\pi)$ 
are always identical due to the 4$\times$4 cluster shape we take. 
%
\bibitem{Ohta94PRB} 
Y. Ohta, K. Tsusui, W. Koshibae, and S. Maekawa: \jo{\PRB}{50}{1994}{13594}. 
%
\bibitem{KurokiPrivate} K. Kuroki: private communications. 
%
\bibitem{Tonegawa87JPSJ} T. Tonegawa and I. Harada: \jo{\JPSJ}{56}{1987}{2153}. 
%
\bibitem{Coldeay96JPCM}
R Coldeay, D A Tennanty, R A Cowleyy, D F McMorrowz, B Dornerx, and
Z Tylczynski: J. Phys.: Condens. Matter {\bf 8} (1996) 7473. 
%
\end{thebibliography}
\end{document}